\documentclass[aps,pra,twocolumn,letterpaper,groupedaddress,10pt]{revtex4-1}
\usepackage{amssymb,amsthm,amsmath,amsfonts}
\usepackage{graphicx,ulem,mathptmx,enumerate,lipsum}
\usepackage[pdftex,dvipsnames,usenames]{xcolor}
\usepackage[colorlinks=true,urlcolor=blue,citecolor=blue,linkcolor=blue]{hyperref}

\begin{document}

\title{Experimental Reconstruction of Entanglement Quasiprobabilities}

\author{J. Sperling}
	\email{jan.sperling@uni-paderborn.de}
	\affiliation{Integrated Quantum Optics Group, Applied Physics, University of Paderborn, 33098 Paderborn, Germany}
\author{E. Meyer-Scott}
	\affiliation{Integrated Quantum Optics Group, Applied Physics, University of Paderborn, 33098 Paderborn, Germany}
\author{S. Barkhofen}
	\affiliation{Integrated Quantum Optics Group, Applied Physics, University of Paderborn, 33098 Paderborn, Germany}
\author{B. Brecht}
	\affiliation{Integrated Quantum Optics Group, Applied Physics, University of Paderborn, 33098 Paderborn, Germany}
\author{C. Silberhorn}
	\affiliation{Integrated Quantum Optics Group, Applied Physics, University of Paderborn, 33098 Paderborn, Germany}

\date{\today}

\begin{abstract}
	We report on the first experimental reconstruction of an entanglement quasiprobability.
	In contrast to related techniques, the negativities in our distributions are a necessary and sufficient identifier of separability and entanglement and enable a full characterization of the quantum state.
	A reconstruction algorithm is developed, a polarization Bell state is prepared, and its entanglement is certified based on the reconstructed entanglement quasiprobabilities, with a high significance and without correcting for imperfections.
\end{abstract}

\maketitle

\paragraph*{Introduction.---}\hspace*{-2.5ex}
	Entanglement is a key feature of any composite quantum system, and it is inconsistent with our classical understanding of correlations.
	For these reasons, this quantum phenomenon has been discussed as a prime example for studying the remarkable features of quantum physics since its discovery  \cite{S35,S36}.
	The implications of the existence of entanglement have been objected to in the seminal EPR paper \cite{ERP35}.
	Later, Bell formulated his famous inequality allowing us to probe for entanglement \cite{B64}.
	Subsequently, based on this inequality, entanglement has been experimentally demonstrated to be a vital part of nature \cite{AGR81}.
	Nowadays, entanglement has become an undisputed property of quantum systems and, in addition, is a versatile resource for realizing, for example, quantum computation and communication tasks beyond classical limitations \cite{NC00,HHHH09}, rendering its verification an essential tool for the function of upcoming quantum technologies.
	Thus, experimentally certifying entanglement is one of the main challenges for realizing such applications \cite{HHHH09,GT09}.

	Mostly independently from the development of the entanglement theory, the notion of quasiprobabilities was devised by Wigner \cite{W32} and others; see Ref. \cite{SW18a} for a thorough introduction.
	In particular, the nonclassicality in a single optical mode can be visualized through negativities in this distribution which cannot occur for classical light.
	For this reason, quantum-optical quasiprobabilities became arguably the most essential and widely applied tool for an intuitive characterization of quantum light in experiments; see Ref. \cite{HSRHMSS16} for a recent implementation.
	Generalizations of the Wigner function can be used to identify nonclassical multimode radiation fields, such as implemented in Ref. \cite{DEWKDKCM13}.
	However, when excluding trivial scenarios (see, e.g., Ref. \cite{DMWS06} for an exception), the negativities in this distribution do not allow for discerning, for instance, single-mode quantum effects from entanglement.

	Still, negativities in quasiprobabilities cover a wide range of applications to describe how quantum effects overcome classical limitations.
	For instance, they can be used to determine the usefulness of a state for quantum information science \cite{F11,VFGE12}.
	Consequently, the study of quasiprobabilities is an active field of current research; see, e.g., the recent Refs. \cite{Z16,SCK17,RHLSJLLL18}.

	To bridge the gap between quasiprobabilities and entanglement, the notion of an entanglement quasiprobability (EQP) distribution has been introduced in theory \cite{STV98,VT99}.
	This approach goes beyond the best approximation to a separable state \cite{LS98,KL01} as it allows not only for convex mixtures but general linear combinations of separable states to expand entangled states, defining EQPs.
	However, the mathematical construction of such distributions is rather complex \cite{SV09b,SW18a}.
	Still, the benefit of EQPs is that negativities in them allow for a necessary and sufficient identification of entanglement.
	Moreover, EQPs apply to discrete- and continuous-variable systems as well as in the multimode scenario beyond bipartite systems \cite{SV12,SW18a}, enabling the theoretical characterization of a manifold of differently entangled states.
	Despite the theoretically predicted advantages of EQPs, to date, EQPs have not been reconstructed in any experiment.

	In this Letter, we report on the experimental reconstruction of EQPs.
	For this proof-of-concept demonstration, we develop the required reconstruction algorithm and generate entangled photon pairs.
	From the correlation measurement of the polarization of the photons, we then directly obtain the EQPs.
	Negativities in that distribution characterize the entanglement of the probed state.
	For comparison, a separable state is prepared and analyzed as well.
	Therefore, our intuitively accessible method to visualize entanglement elevates the versatile notion of EQPs to a practical tool to characterize sources of quantum light in experiments.

\paragraph*{Theory of EQPs.---}\hspace*{-2.5ex}
	By definition \cite{W89}, a bipartite mixed separable state can be written in the form
	\begin{align}
		\label{eq:SepState}
		\hat\rho=\sum_{a,b}P(a,b)|a,b\rangle\langle a,b|,
	\end{align}
	using the normalized tensor-product vectors $|a,b\rangle=|a\rangle\otimes|b\rangle$.
	Therein, $P$ is a classical, i.e., non-negative, probability distribution, $P\geq0$.
	A state is entangled (likewise, inseparable) iff it cannot be written according to Eq. \eqref{eq:SepState}.
	However, if we allow for $P$ to be a quasiprobability distribution which can take negative values, $P\ngeq0$, any entangled state can be expanded as shown in Eq. \eqref{eq:SepState} as well \cite{STV98,VT99}; see also Refs. \cite{SV09b,SW18a}.

	A general decomposition of a state in terms of pure separable ones is not unique, and the challenge is to find an optimal representation \cite{SV09b}.
	Ignoring the property of separability, we find a similar problem for the general expansion of a single-mode density operator in terms of pure states \cite{SW18a}.
	There, the optimal expansion is found through the spectral decomposition which allows us to write any state in terms of its eigenstates and nonnegative eigenvalues.
	Thus, the eigenvalue equation for the density operator has to be solved.
	An unphysical operator, on the other hand, has negative contributions in its spectral decomposition.
	A similar approach to the spectral decomposition can be found for composite system when including the property of separability, leading to EQPs.

	For this reason, the so-called separability eigenvalue equations have been developed \cite{SV09b,SW18a},
	\begin{align}
		\label{eq:SEEs}
		\hat\rho_b|a\rangle=g|a\rangle
		\quad\text{and}\quad
		\hat\rho_a|b\rangle=g|b\rangle,
	\end{align}
	where $\hat\rho_b=\mathrm{tr}_B[\hat\rho(\hat 1_A\otimes|b\rangle\langle b|)]$ and $\hat\rho_a=\mathrm{tr}_A[\hat\rho(|a\rangle\langle a|\otimes\hat 1_B)]$, which can be further generalized to multipartite system \cite{SW18a}.
	The different vectors $|a_i,b_i\rangle$ and values $g_i$ that solve Eq. \eqref{eq:SEEs} refer to as separability eigenvectors and separability eigenvalues, respectively.
	Here, $i$ describes, in general, a multi-index that lists the individual solutions.
	The approach of coupled eigenvalue equations in Eq. \eqref{eq:SEEs} also enables the construction of entanglement witnesses \cite{SV09a,SV13}.
	Furthermore, the solutions allow us to expand the state as $\hat\rho=\sum_{i}P(a_i,b_i)|a_i,b_i\rangle\langle a_i,b_i|$.
	In particular, all values of the EQP, $\vec p=[P(a_i,b_i)]_i$, are obtained from the solution of the linear equation \cite{SV09b,SW18a}
	\begin{align}
		\label{eq:LinPart}
		\mathbf{G} \vec p=\vec g,
	\end{align}
	using the Gram-Schmidt matrix $\mathbf{G}=(|\langle a_i,b_i|a_j,b_j\rangle|^2)_{i,j}$ and the vector $\vec g=(g_j)_j$ of separability eigenvalues.

	Most importantly, it has been proven that the presented approach yields a nonnegative $\vec p$ [i.e., $\forall i: P(a_i,b_i)\geq0$] for any separable state and includes negative entries for any inseparable state \cite{SW18a}.
	In this sense, this technique is optimal, and $\vec p\ngeq0$ is our necessary and sufficient entanglement criterion.
	In analogy to the spectral decomposition, the expansion of states in composite systems [cf. Eq. \eqref{eq:SepState}] can be achieved using the separability eigenvectors [Eq. \eqref{eq:SEEs}] and the EQP obtained from the separability eigenvalues [Eq. \eqref{eq:LinPart}].

	For our purpose, we consider a two-qubit state.
	Any such state can be recast into the so-called standard form to analyze correlations using local transformations only \cite{LMO06}.
	It reads
	\begin{align}
		\label{eq:Normal}
		\hat\rho^\mathrm{(std)}
		=\frac{\rho_0}{4}\hat\sigma_0{\otimes}\hat\sigma_0
		+\frac{\rho_x}{4}\hat\sigma_x{\otimes}\hat\sigma_x
		+\frac{\rho_y}{4}\hat\sigma_y{\otimes}\hat\sigma_y
		+\frac{\rho_z}{4}\hat\sigma_z{\otimes}\hat\sigma_z,
	\end{align}
	where $\hat\sigma_x$, $\hat\sigma_y$, and $\hat\sigma_z$ are the Pauli matrices, completed with the identity $\hat\sigma_0$;
	note that $\rho_0=1$ is the normalization.

\begin{figure}
	\includegraphics[width=\columnwidth]{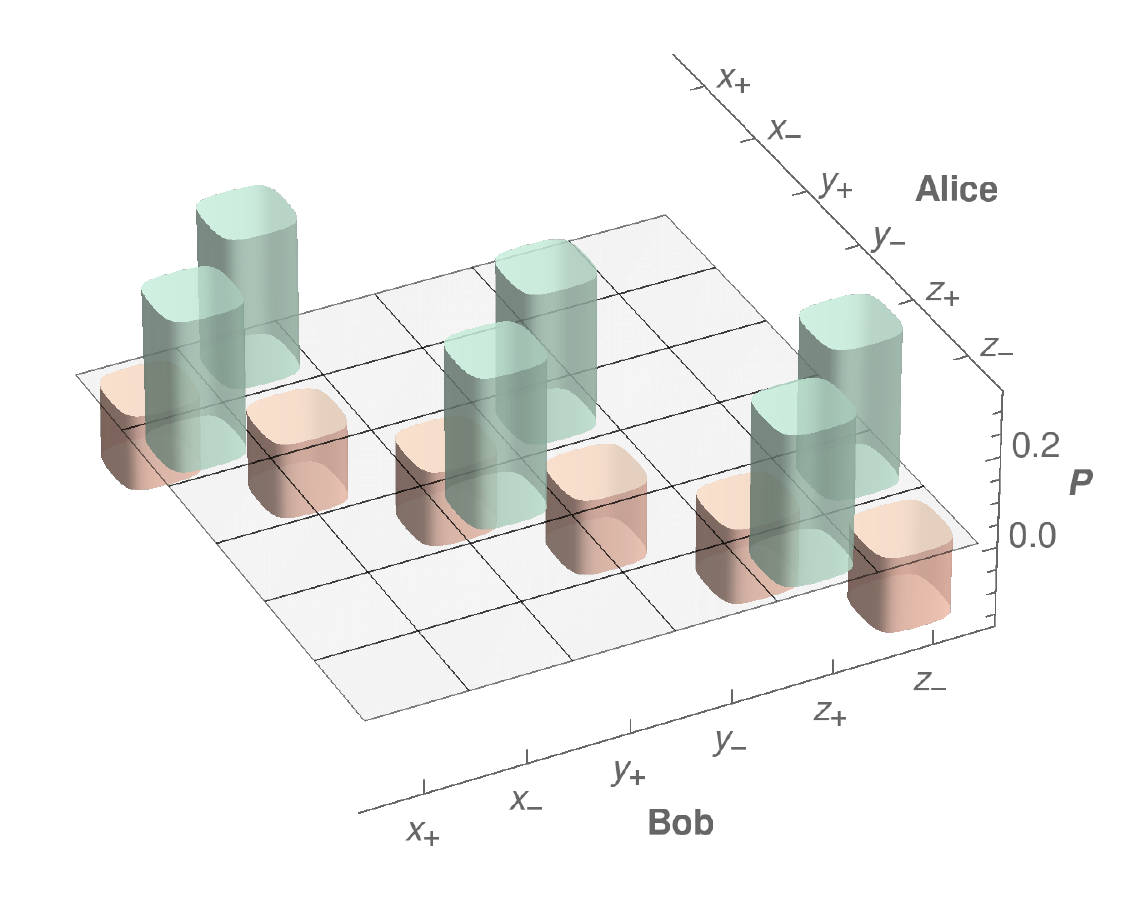}
	\caption{
		Ideal EQP of the target state $|\psi\rangle=(|H,V\rangle-|V,H\rangle)/\sqrt2$, which is given by $\rho_x=\rho_y=\rho_z=-1=-\rho_0$, cf. Eqs. \eqref{eq:Normal} and \eqref{eq:StdDecomp}.
		The EQP $P$ over Alice's and Bob's subsystems, which is determined through the eigenstates of Pauli operators, takes negative values to account for the presence of quantum entanglement, $P\ngeq0$.
	}\label{fig:1}
\end{figure}

	The exact solution of the separability eigenvalue equations \eqref{eq:SEEs} for states in the standard form \eqref{eq:Normal} has been formulated and its EQP was subsequently determined \cite{SW18a}.
	In particular, it was shown that the state can be written as
	\begin{align}
		\label{eq:StdDecomp}
		\hat\rho^\mathrm{(std)}=\sum_{\substack{
			w\in\{x,y,z\}
			\\
			\alpha,\beta\in\{+,-\}
		}}
		P^\mathrm{(std)}(w_\alpha,w_\beta)|w_\alpha,w_\beta\rangle\langle w_\alpha,w_\beta|,
	\end{align}
	where the exact form of $P^\mathrm{(std)}$ can be additionally found in the Supplemental Material (SM) \cite{supplement}.
	The separability eigenvectors used in Eq. \eqref{eq:StdDecomp} are $|w_\pm\rangle$ for both Alice's and Bob's subsystem, which are the eigenvectors to the corresponding Pauli matrices, $\hat\sigma_w$  for $w\in\{x,y,z\}$.
	For example, the EQP for our target state---the polarization Bell state $|\psi\rangle=(|H,V\rangle-|V,H\rangle)/\sqrt2$---is shown in Fig. \ref{fig:1}.
	The distinct negativities directly visualize the entanglement of this state.

\paragraph*{Reconstruction algorithm for EQPs.---}\hspace*{-2.5ex}
	Let us now outline how to obtain the EQPs.
	For a detailed step-by-step description of the developed method, we refer to the SM \cite{supplement}.
	The first step of our reconstruction algorithm is to apply local transformations to obtain the standard form.
	Then, we use the known EQP for the standard-form state and apply the inverse transformations to find the EQP for the measured state.
	
	In the first step, we find the local operations to relate the measured state $\hat\rho$ with its corresponding standard form \eqref{eq:Normal},
	\begin{align}
		\label{eq:Trafo}
		\hat\rho^\mathrm{(std)}=(\hat T_A\otimes\hat T_B)^{-1}\hat\rho(\hat T_A^\dag\otimes\hat T_B^\dag)^{-1},
	\end{align}
	where $\hat T_A\otimes\hat T_B$ are local invertible transformations that can be constructed \cite{supplement} by adapting methods from Refs. \cite{LMO06,SCS99}.
	The transformations consist of rotations and Lorentz-type operations which jointly result in the desired standard form.
	It is important to stress that these local transformations do not affect the property of a state of being entangled or not \cite{SV11}.

	In the second step, the known solutions of the separability eigenvalue equations of the states $\hat\rho^\mathrm{(std)}$ are applied.
	Consequently, the EQPs can be directly obtained using the local transformations constructed by our algorithm.
	Equating Eqs. \eqref{eq:StdDecomp} and \eqref{eq:Trafo}, we find
	\begin{align}
		\label{eq:StateExpansion}
		\hat\rho=&\sum_{\substack{ \overline w\in\{\overline x,\overline y,\overline z\} \\ \alpha,\beta\in\{+,-\} }}
		P(\overline w_\alpha,\overline w_\beta)|\overline w_\alpha,\overline w_\beta\rangle\langle \overline w_\alpha,\overline w_\beta|,
	\end{align}
	using the locally transformed distribution $P(\overline w_\alpha,\overline w_\beta)=\langle w_\alpha|\hat T_A^\dag\hat T_A|w_\alpha\rangle\langle w_\beta|\hat T_B^\dag\hat T_B|w_\beta\rangle P^\mathrm{(std)}(w_\alpha,w_\beta)$ together with the normalized and transformed tensor-product states $|\overline w_\alpha,\overline w_\beta\rangle=\hat T_A|w_\alpha\rangle/\langle w_\alpha|\hat T_A^\dag\hat T_A|w_\alpha\rangle^{1/2}\otimes\hat T_B|w_\beta\rangle/\langle w_\beta|\hat T_B^\dag\hat T_B|w_\beta\rangle^{1/2}$.
	Errors are estimated via a standard Monte Carlo approach \cite{supplement}.
	We also emphasize that because of Eq. \eqref{eq:StateExpansion}, both the state and its entanglement are fully characterized through its EQP and the transformed separability eigenstates.
	By contrast, the density operator reconstruction alone does not directly yield the entanglement features.

\paragraph*{Experimental implementation.---}\hspace*{-2.5ex}
	We produce two-photon states via parametric down-conversion in a periodically poled titanyl phosphate waveguide in a Sagnac loop  \cite{KFW06};
	full details on the source can be found in Ref. \cite{MPEQDBS18}.
	The source is pumped bidirectionally with $200\,\mathrm{\mu W}$ pulsed light at $770\,\mathrm{nm}$ wavelength, producing polarization-entangled photon pairs at $1540\,\mathrm{nm}$.
	For our measurements, the source produces $330\,000$ photon pairs per second in each direction, of which $59\,000$ per second are finally detected, with a total detection efficiency for the signal mode of $38\%$ and the idler mode of $47\%$.
	We collect coincidence counts for $36$ polarization measurement settings, where Alice and Bob independently set their polarizers to horizontal, vertical, diagonal, antidiagonal, right circular, and left circular, corresponding to $x$, $y$, and $z$ measurements.
	Collecting data for one second for each setting gives us just over one million total coincidence counts, which we then feed into the EQP reconstruction algorithm.

	We additionally performed a maximum likelihood quantum state reconstruction \cite{JKMW01}, finding an overlap with the targeted polarization-entangled Bell state of $0.9578\pm0.0004$.
	This is consistent with the direct sampling approach used here \cite{supplement} that gives the overlap $0.958\pm0.003$.
	In addition to the entangled state, we prepared a separable state which corresponds to the target state $(|H\rangle+|V\rangle)/\sqrt 2\otimes|H\rangle$.
	To generate this state, we pumped the source in just one direction to produce $|V\rangle\otimes|H\rangle$ photon pairs and then inserted a half-wave plate at $22.5^\circ$ in the signal (Alice's) arm to rotate the first photon to the desired superposition state.

\begin{figure}
	\includegraphics[width=\columnwidth]{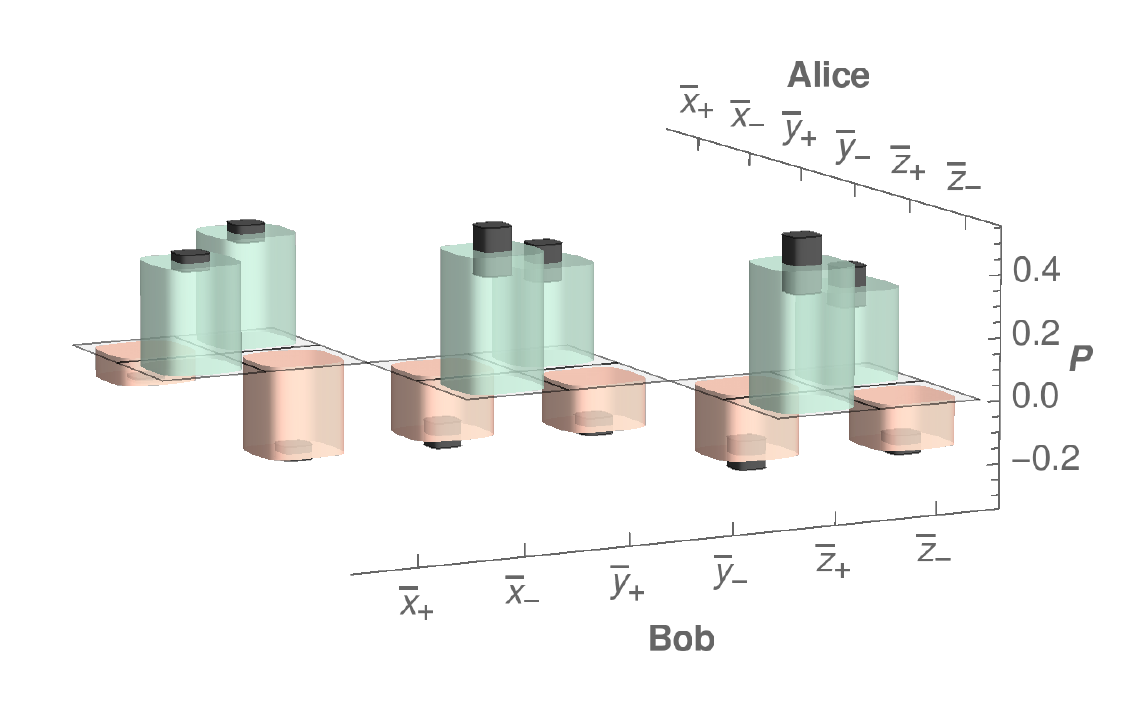}
	\caption{
		Reconstructed EQP including error bars.
		Significant negativities, up to 13 standard deviations, directly certify the quantum correlation between Alice's and Bob's subsystem.
		A direct comparison with the ideal case is additionally provided in the SM \cite{supplement}.
	}\label{fig:2}
\end{figure}

\paragraph*{Results.---}\hspace*{-2.5ex}
	We apply our developed reconstruction algorithm to our data.
	The resulting EQP is depicted in Fig. \ref{fig:2}.
	The negativities demonstrate the entanglement of the generated state with a significance of $13$ standard deviations.
	This proves that EQPs are not only a mathematical concept, but a useful tool to experimentally characterize the quantumness of correlations.

	The EQP in Fig. \ref{fig:2} is in good agreement with the ideal case;
	the asymmetries in Fig. \ref{fig:2} when compared to Fig. \ref{fig:1} are a result of the rescaling due to the local transformation [see the description of Eq. \eqref{eq:StateExpansion}].
	In this context, let us stress that our reconstruction does not correct for any imperfections.
	That is, the experimentally obtained EQP (Fig. \ref{fig:2}) includes all impurities of the setup and still verifies the entanglement with high significance.
	This further shows that EQPs can be used to assess the high quality of our setup as a reliable source for entangled states.
	Additional properties resulting from our analysis are provided in the SM \cite{supplement}.
	For instance, the reconstruction density operator $\hat\rho$ via Eq. \eqref{eq:StateExpansion}, also using the reconstructed states $|\overline w_\alpha,\overline w_\beta\rangle$, is confirmed by the direct sampling approach.

\begin{figure}
	\includegraphics[width=\columnwidth]{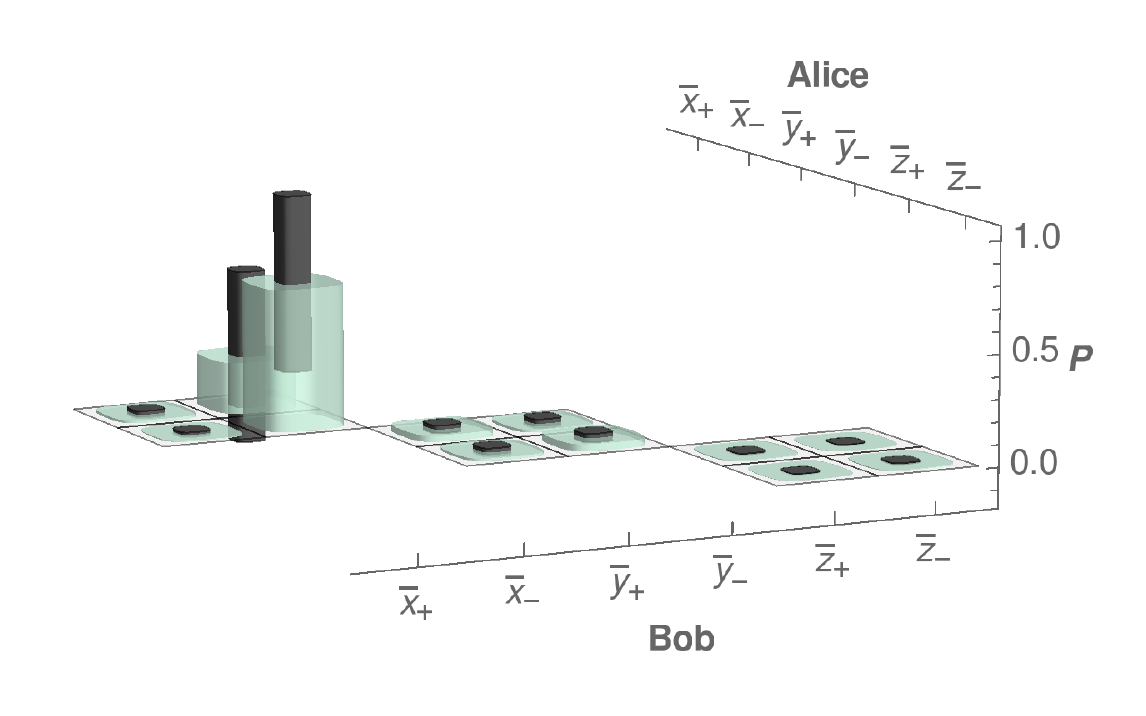}
	\caption{
		Reconstruction for a separable target state $(|H\rangle+|V\rangle)/\sqrt 2\otimes|H\rangle$.
		Without requiring additional testing, it is directly evident that the produced state is separable, $P\geq0$.
		The comparably large error bars are a result of the involved matrix inversion [Eq. \eqref{eq:Trafo}].
	}\label{fig:3}
\end{figure}

	Furthermore, to challenge our reconstruction procedure, we also analyzed a separable state.
	In Fig. \ref{fig:3}, the resulting EQP is shown which does not include any negative contribution.
	It is worth recalling that the employed transformations can include rotations in the $x$-$y$-$z$ space.
	Interestingly, the nonnegativity is a direct proof of the state's separability, which would require an informationally complete number of other entanglement tests to rule out entanglement for any sufficient criterion.
	The comparably large error bars are a result of the fact that the measured correlation matrix is close to being noninvertible, resulting in high fluctuations when determining the local transformation $\hat T_A\otimes \hat T_B$ [cf. Eq. \eqref{eq:Trafo}].
	A similar behavior is found for other inversion problems when performing reconstruction tasks; see the discussions, e.g., in Refs. \cite{L04,SSG09}.

\paragraph*{Discussion.---}\hspace*{-2.5ex}
	In contrast to other well-known quasiprobability distributions, EQPs are developed to specifically determine entanglement.
	In comparison, the Wigner function of an entangled two-mode squeezed-vacuum state is a nonnegative Gaussian distribution, while the Wigner function of an uncorrelated product state of a single photon and vacuum includes negativities.
	As such, commonly reconstructed distributions are inconclusive when assessing the state's entanglement.
	By contrast, the significant negativities in our experimentally reconstructed EQP in Fig. \ref{fig:2} exclusively and unambiguously certify the generated entanglement in a direct manner.

	The partial transposition (PT) criterion \cite{P96,HHH96} is another approach to uncover entanglement in a two-qubit system.
	In fact, this yields a useful benchmark to independently validate our results \cite{supplement}.
	However, the PT fails to be a necessary and sufficient criterion for general systems.
	Beyond such a limitation, EQPs have been theoretically proven to apply to arbitrary composite systems, including continuous-variable \cite{SV12} and multimode states \cite{SW18a}.
	Furthermore, a generalization of the standard form to qudit systems, which is required for such a purpose, was already discussed in Ref. \cite{LMO06} and enables the generalization of our developed approach.

	The general technique to obtain EQPs [cf. Eqs. \eqref{eq:SEEs} and \eqref{eq:LinPart}] is based on the separability eigenvalue equations [Eq. \eqref{eq:SEEs}].
	As mentioned previously, this method also allows for the formulation of entanglement witnesses \cite{SV09a,SV13}, which, for example, enable the experimental detection of complex forms of multipartite entanglement in continuous-variable systems \cite{GSVCRTF15,GSVCRTF16}.
	Recently, a numerical approach to solve those equations in cases where an exact solution is unknown has been devised as well \cite{GVS18}.
	However, identifying separability through entanglement witnesses requires testing a large number of them, in contrast to the direct criterion of the nonnegativity of EQPs.
	In addition to witnesses, even the entanglement dynamics of a composite system is accessible through a time-dependent version of the presented eigenvalue approach \cite{SW17b}.

	The inclusion of the notion of EQPs shows the universality of the method of separability eigenvalue equations as well as its robustness.
	In this context, the usefulness of EQPs in theoretical studies has also been exemplified for NOON states propagating in atmospheric loss channels \cite{BSV17}, two-mode squeezed states under the influence of dephasing \cite{SV12}, and multipartite bound-entangled states (inaccessible with the PT criterion) \cite{SW18a}.
	Here, we complement this versatile theory with its experimental realization.

\paragraph*{Summary.---}\hspace*{-2.5ex}
	We developed a reconstruction algorithm that renders it possible to experimentally apply the notion of entanglement quasiprobabilities (EQPs), providing the missing link between the theory of EQPs and its experimental implementation.
	With this approach, we verified the entanglement of a two-mode polarization state|the basis for many quantum information protocols|which is not possible with other well-known quasiproabilities but was theoretically predicted two decades ago \cite{STV98}.
	Similarly to the spectral decomposition of physical density operators, the solutions of the separability eigenvalue equations enable the expansion of any separable state in terms of a classical joint distribution.
	We confirmed this by producing and probing a separable state, which additionally overcomes the so-called separability problem that aims at certifying that a state is indeed separable.
	Conversely, entanglement is unambiguously verified through negativities in the EQP distribution.
	Here, this versatile, necessary, and sufficient theory was used to assess separability and inseparability experimentally, which has not been done before.
	By developing a reconstruction technique for EQPs, we were able to certify and visualize entanglement through the negativities in the EQP with high statistical significance.
	This was achieved without correcting for experimental imperfections.
	Therefore, we implemented a previously inaccessible and intuitive method to experimentally uncover entanglement for future applications in quantum technologies.

\paragraph*{Acknowledgments.---}\hspace*{-2.5ex}
	J. S. is grateful to Alexander Reusch for enlightening discussions.
	E. M.-S. acknowledges funding from the Natural Sciences and Engineering Research Council of Canada (NSERC).
	The Integrated Quantum Optics group acknowledges financial support from European Commission with the ERC project QuPoPCoRN (No. 725366) and from the Gottfried Wilhelm Leibniz-Preis (Grant No.  SI1115/3-1).

\onecolumngrid
\appendix*
\section{Supplemental Material}
\twocolumngrid

	Here, we provide all details for the reconstruction of the entanglement quasiprobability (EQP) and show in a step-by-step manner how the reconstruction applies to our data for the entangled state.

\subsection{Notations for polarization qubits}\label{app:Qubit}

	For a single qubit, we chose the computational basis to resemble the horizontal and vertical polarization, $\{|H\rangle,|V\rangle\}$.
	In this basis, we can expand the vectors for the diagonal and anti-diagonal polarization as $|D\rangle=(|H\rangle+|V\rangle)/\sqrt2$ and $|A\rangle=(|H\rangle-|V\rangle)/\sqrt2$, respectively.
	For the circular polarization, we also get the vector for right, $|R\rangle=(|H\rangle+i|V\rangle)/\sqrt2$, and left, $|L\rangle=(|H\rangle-i|V\rangle)/\sqrt2$.
	\begin{subequations}
		\label{eq:PauliOp}
	Consequently, the Pauli matrices can be written for our specific purposes as
	\begin{align}
		\hat\sigma_x
		=|H\rangle\langle H|-|V\rangle\langle V|
		=\begin{bmatrix}1&0\\0&-1\end{bmatrix},
	\end{align}
	\begin{align}
		\hat\sigma_y
		=|D\rangle\langle D|-|A\rangle\langle A|
		=\begin{bmatrix}0&1\\1&0\end{bmatrix},
	\end{align}
	and
	\begin{align}
		\hat\sigma_z
		=|R\rangle\langle R|-|L\rangle\langle L|
		=\begin{bmatrix}0&-i\\i&0\end{bmatrix}.
	\end{align}
	\end{subequations}
	It is also worth mentioning that the identity reads $\hat\sigma_0=\left[\begin{smallmatrix}1&0\\0&1\end{smallmatrix}\right]$.
	Considering a bipartite system, all notations can be canonically extended.
	For example, our bipartite computational basis reads $\{|H,H\rangle,|H,V\rangle,|V,H\rangle,|V,V\rangle\}$.

\subsection{Solution of the standard form}

	The separability eigenvalue equations for an operator $\hat\rho^\mathrm{(std)}$ in the standard form, defined through the set of parameters $[\rho_{0},\rho_{x},\rho_{y},\rho_{z}]$, have been solved in Ref. \cite{SW18a}.
	In particular, the EQP that has been obtained reads
\begin{widetext}
	\begin{align}
		\label{eq:PentSol}
		\mathbf{P}^\mathrm{(std)}=\begin{bmatrix}
			\frac{q}{12}+\frac{|\rho_{x}|+\rho_{x}}{4} & \frac{q}{12}+\frac{|\rho_{x}|-\rho_{x}}{4} & 0 & 0 & 0 & 0
			\\
			\frac{q}{12}+\frac{|\rho_{x}|-\rho_{x}}{4} & \frac{q}{12}+\frac{|\rho_{x}|+\rho_{x}}{4} & 0 & 0 & 0 & 0
			\\
			0 & 0 & \frac{q}{12}+\frac{|\rho_{y}|+\rho_{y}}{4} & \frac{q}{12}+\frac{|\rho_{y}|-\rho_{y}}{4} & 0 & 0
			\\
			0 & 0 & \frac{q}{12}+\frac{|\rho_{y}|-\rho_{y}}{4} & \frac{q}{12}+\frac{|\rho_{y}|+\rho_{y}}{4} & 0 & 0
			\\
			0 & 0 & 0 & 0 & \frac{q}{12}+\frac{|\rho_{z}|+\rho_{z}}{4} & \frac{q}{12}+\frac{|\rho_{z}|-\rho_{z}}{4}
			\\
			0 & 0 & 0 & 0 & \frac{q}{12}+\frac{|\rho_{z}|-\rho_{z}}{4} & \frac{q}{12}+\frac{|\rho_{z}|+\rho_{z}}{4}
		\end{bmatrix},
	\end{align}
\end{widetext}
	where $q=\rho_{0}-|\rho_{x}|-|\rho_{y}|-|\rho_{z}|$ and the entries of $\mathbf{P}^\mathrm{(std)}$ define $P^\mathrm{(std)}(a,b)$.
	Therein, the rows and columns refer to Alice's ($a$) and Bob's ($b$) subsystem, respectively, and are ordered according the eigenvectors of the Pauli operators [Eq. \eqref{eq:PauliOp}] to the eigenvalues $\pm1$, i.e., $a,b\in\{x_+,x_-,y_+,y_-,z_+,z_-\}$.
	Note that the zero entries do not correspond to vectors that solve the underlying separability eigenvalue equations and are, therefore, set to zero to enable the matrix representation \eqref{eq:PentSol}.

\subsection{Sampling the correlation matrices}\label{app:DensityOp}

\begin{figure}
	\includegraphics[width=.75\columnwidth]{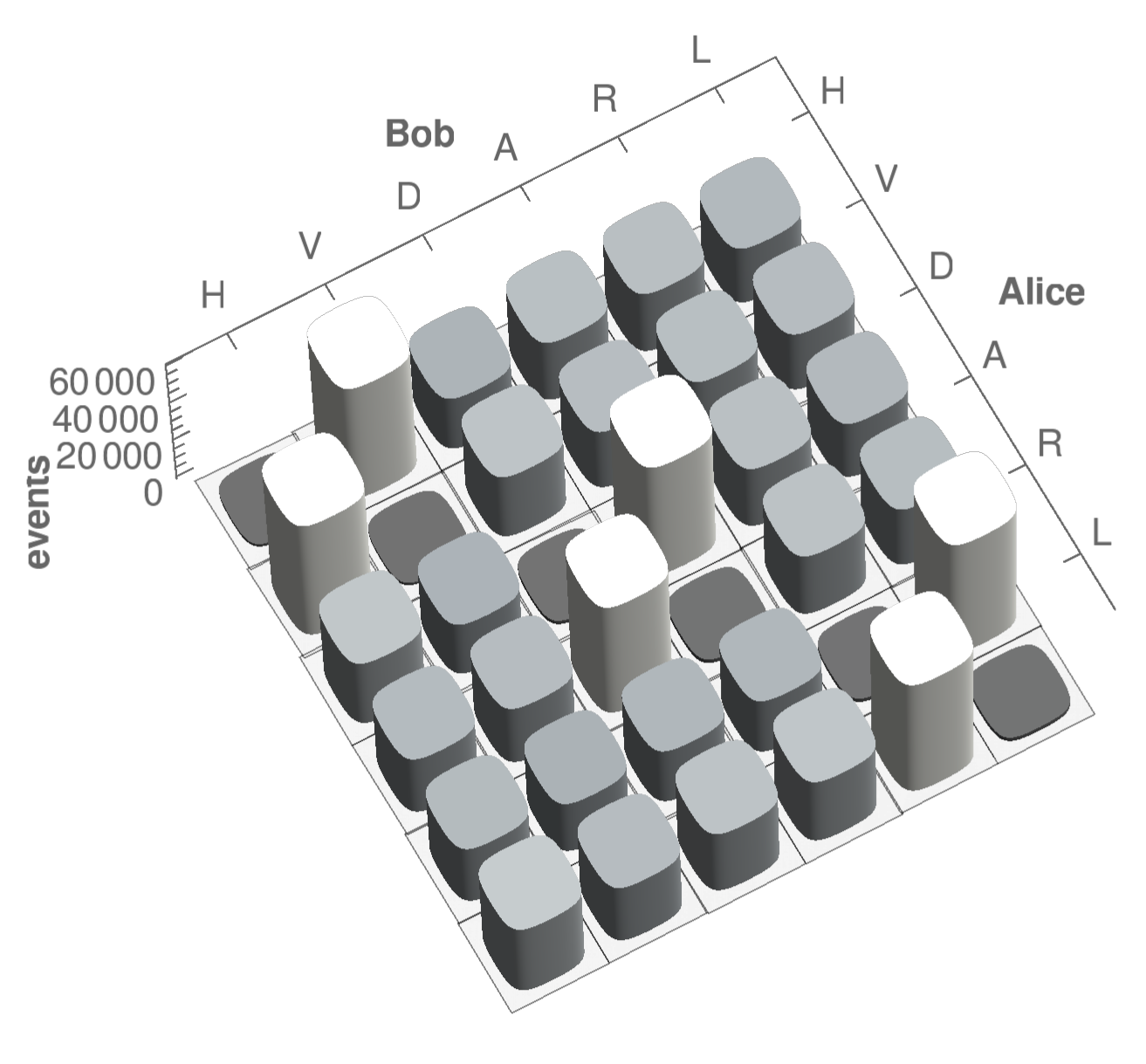}
	\caption{
		Raw data from polarization measurements.
	}\label{fig:Supp1}
\end{figure}

	Our measurement of the different polarization components yields the coincidence events $E(s,t)$, where $s,t\in\{H,V,D,A,R,L\}$ describe the outcomes.
	For convenience, the coincidences are collected in the matrix $\mathbf{E}=[E(s,t)]_{s,t}$.
	The corresponding data are depicted in Fig. \ref{fig:Supp1}.

\begin{figure*}
	\includegraphics[width=.95\textwidth]{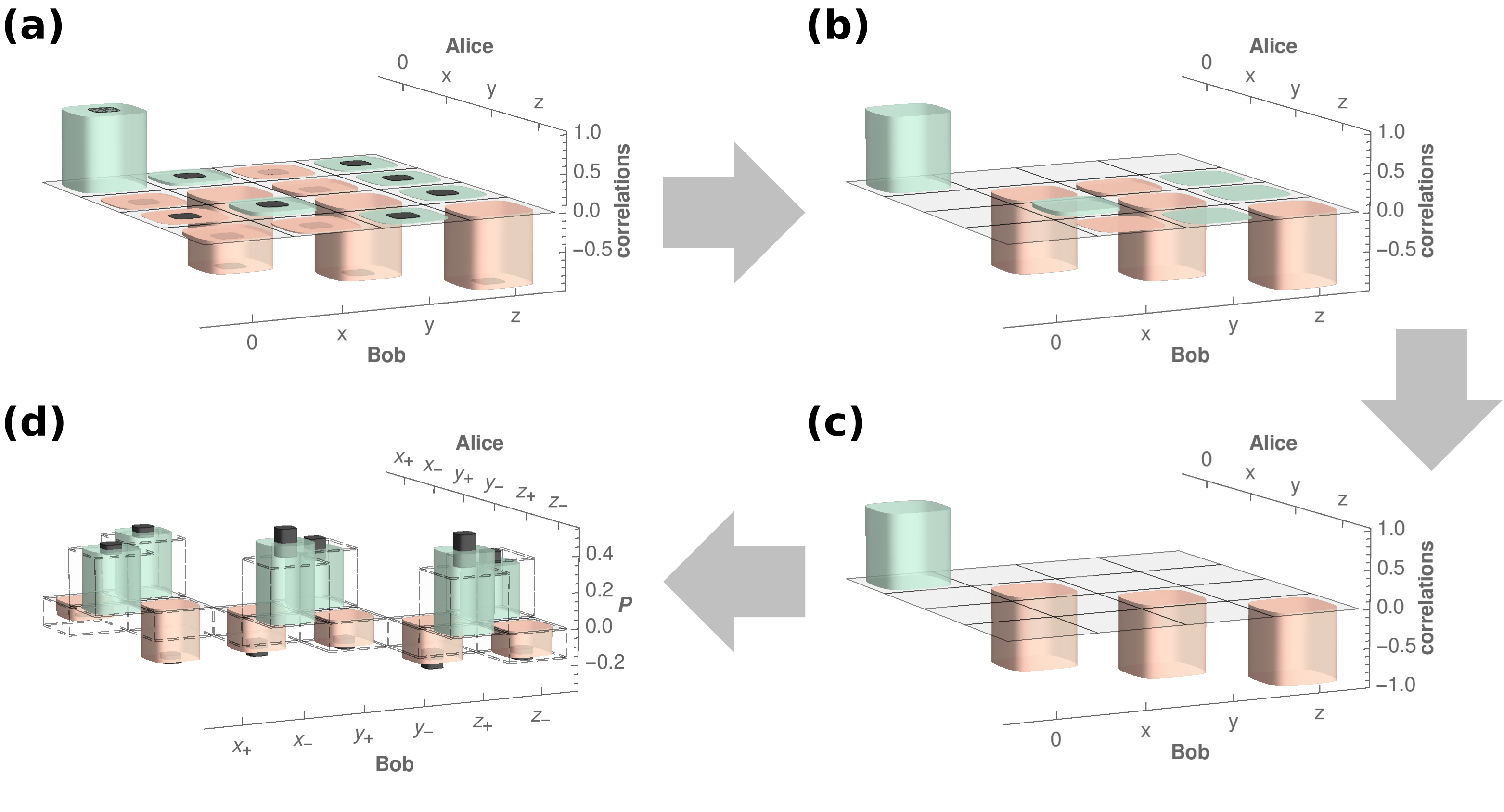}
	\caption{
		(a) Sampled correlation matrix with the entries $\langle\hat\sigma_k\otimes\hat\sigma_l\rangle$ for $k,l\in\{0,x,y,z\}$.
		(b) Correlation matrix after local boost operation to get $\langle\hat\sigma_k\otimes\hat\sigma_0\rangle=0=\langle\hat\sigma_0\otimes\hat\sigma_l\rangle$.
		(c) Correlation matrix after local rotations for diagonalization.
		(d) Reconstructed EQP including one standard deviation error bar (black cylinders).
		The dashed boxes show the EQP of the ideal target state for comparison.
	}\label{fig:Supp2}
\end{figure*}

	To obtain the correlations $\langle \hat \sigma_k\otimes\hat\sigma_l\rangle$, a sampling approach is devised.
	It is again convenient to define a matrix,
	\begin{align}
		\label{eq:ExactEQP}
		\mathbf{S}=\begin{bmatrix}
			1 & 1 & 1 & 1 & 1 & 1
			\\
			1 & -1 & 0 & 0 & 0 & 0
			\\
			0 & 0 & 1 & -1 & 0 & 0
			\\
			0 & 0 & 0 & 0 & 1 & -1
		\end{bmatrix},
	\end{align}
	in which the rows relate to the Pauli matrices in the order $\{0,x,y,z,\}$, and the columns are ordered according to $\{H,V,D,A,R,L\}$.
	Then, we simply get the the correlations and their error estimates in the following compact form:
	\begin{align}
		\mathbf{C}=&\Big[
			\langle \hat \sigma_k\otimes\hat\sigma_l\rangle
		\Big]_{k,l\in\{0,x,y,z\}}
		=(\mathbf{S}\mathbf{E}\mathbf{S}^\mathrm{T})/(|\mathbf{S}|\mathbf{E}|\mathbf{S}|^\mathrm{T})
		\text{ and}
		\\\nonumber
		\Delta\mathbf{C}=&\sqrt{\big(
			(\mathbf{S}^2\mathbf{E}\mathbf{S}^{2\mathrm{T}})/(|\mathbf{S}|\mathbf{E}|\mathbf{S}|^\mathrm{T})-\mathbf{C}*\mathbf{C}
		\big) /\big(
			|\mathbf{S}|\mathbf{E}|\mathbf{S}|^\mathrm{T}-\mathbf{1}
		\big)},
	\end{align}
	where the operations ``$|\,\,|$'', ``$*$'', ``$/$'', and ``${\,\,}^2$'' act entry-wise and $\mathbf 1$ denotes a matrix with all entries being one.
	Note that $\mathbf{S^2}=|\mathbf{S}|$.
	The resulting correlation matrix is shown in Fig. \ref{fig:Supp2}(a).
	Furthermore, the reconstructed state is obtained via $\hat\rho=\sum_{k,l\in\{0,x,y,z\}}C_{k,l}\hat\sigma_k\otimes\hat\sigma_l/4$.
	Except for the compact notation, this part of the reconstruction is a standard sampling approaches, as widely used in quantum optics.

\subsection{Generalized singular value decomposition}

	In order to get the standard form of the correlation matrix, $\mathbf{C}^\mathrm{(std)}=\mathrm{diag}[C^\mathrm{(std)}_{j,j}]_{j\in\{0,x,y,z\}}$ (corresponding to $\hat\rho^\mathrm{(std)}$), appropriate transformations have to be performed on the current correlation matrix $\mathbf{C}=[C_{j,k}]_{j,k\in\{0,x,y,z\}}$ (corresponding to $\hat\rho$).
	For this purpose, a generalization of the singular value decomposition is performed,
	\begin{align}
		\label{eq:TrafoToStd}
		\mathbf{C}^\mathrm{(std)}=
		(\mathbf{R}_A\mathbf{L}_A)
		\mathbf{C}
		(\mathbf{L}_B^T\mathbf{R}_B^T),
	\end{align}
	where $\mathbf{R}_{A}$ and $\mathbf{R}_{B}$ are rotations and $\mathbf{L}_{A}$ and $\mathbf{L}_{B}$ resemble Lorentz-type transformations \cite{LMO06}.

\subsubsection{Local boost operations}

	Let us have a look at the Lorentz-like operations, which also refer to as boost operations.
	How such operations can be used to find the standard form was described in Ref. \cite{LMO06}, and a general construction approach is provided in Ref. \cite{SCS99}.
	The aim of this transformation is to eliminate the local components $[C_{0,x},C_{0,y},C_{0,z}]$ and $[C_{x,0},C_{y,0},C_{z,0}]$, cf. Fig. \ref{fig:Supp2}(b).

	The class of operations $\mathbf{L}$ is defined through satisfying the relation $\mathbf{L}\mathbf{\eta}\mathbf{L}^T=\mathbf{\eta}$, with the metric $\mathbf{\eta}=\mathrm{diag}[1,-1,-1,-1]$ and the latter three components being the spatial part.
	As an example, this transforms a (single qubit) vector $[\gamma_0,\gamma_x,\gamma_y,\gamma_z]$, representing a state $\hat\gamma=\sum_{j\in\{0,x,y,z\}}\gamma_j\hat\sigma_j$, as applying a matrix
	\begin{align}
		\mathbf{L}=\begin{bmatrix}
			\cosh(r) & \sinh(r) w^T
			\\
			\sinh(r) w & \cosh(r) \mathbf{P} +\overline{\mathbf{P}}
		\end{bmatrix},
	\end{align}
	with $w^Tw=1$, $\mathbf{P}=ww^T$, and $\overline{\mathbf{P}}=\mathrm{diag}[1,1,1]-ww^T$.
	For example, this acts on the density operator as
	\begin{align}
	\begin{aligned}
		\hat T^\prime\hat\gamma\hat T^{\prime\dag}=
		&\left(\cosh(r)\gamma_0+\sinh(r)\gamma_z\right)\hat\sigma_0
		\\
		&+\gamma_x\hat\sigma_x+\gamma_y\hat\sigma_y
		\\
		&+\left(\sinh(r)\gamma_0+\cosh(r)\gamma_z\right)\hat\sigma_z
	\end{aligned}
	\end{align}
	for the case $w=[0,0,1]^T$ and with $\hat T'=\mathrm{diag}[e^{r/2},e^{-r/2}]$.
	Similarly, a transformation $\mathbf{L}_A\mathbf{C}\mathbf{L}_B^T$ corresponds to a separable, bipartite operation $(\hat T'_A\otimes\hat T'_B)\hat\rho(\hat T'_A\otimes\hat T'_B)^\dag$.

	In practice, we compute $\mathbf{C}\eta\mathbf{C}^T$ for Alice's side to get a symmetric matrix $\mathbf{M}=\left[\begin{smallmatrix} \alpha & \beta^T \\ \beta & \gamma \end{smallmatrix}\right]$, where $\gamma=\gamma^T$.
	The requirement that
	\begin{align}
		\mathbf{L}_A\mathbf{M}\mathbf{L}_A^T
		=\begin{bmatrix}
			\cdot & 0^T
			\\
			0 & \cdot
		\end{bmatrix},
	\end{align}
	with ``$\cdot$'' indicating irrelevant entries, yields the following conditions in which $v=\tanh(r)w$:
	$0=\overline{\mathbf{P}}(\beta+\gamma v)$ and $0=\alpha+(1+1/v^Tv)\beta^Tv+v^T\gamma v/v^Tv$.
	The first condition is solved via $v=(\lambda\mathrm{diag}[1,1,1]-\gamma)^{-1}\beta$ by introducing the real valued parameter $\lambda$.
	Note that from $v^Tv=\tanh^2(r)$, we can determine $r\geq0$ and, thus, $w=v/\tanh(r)$.
	Inserting this into the second condition results in an equation for $\lambda$, $0=\alpha+\lambda+\beta^T(\lambda-\gamma)^{-1}\beta$, which we solve numerically.

	Analogously, we obtain $\mathbf{L}_B$ from $\mathbf{M}=\mathbf{C}^T\eta\mathbf{C}$.
	The result of both local transformations is the transition shown from (a) to (b) in Fig. \ref{fig:Supp2}.

\subsubsection{Local rotations}

	Now, we can perform local rotation $\mathbf{R}_{A}$ and $\mathbf{R}_{B}$ on the spatial parts of $\mathbf{L}_A\mathbf{C}\mathbf{L}_B^T$ to get the standard form.
	This corresponds to applying local unitaries, $(\hat U_A\otimes\hat U_B)\hat\rho(\hat U_A\otimes\hat U_B)^\dag$.
	The desired rotations can be obtained from a standard singular-value decomposition.

	Note that reflections do not belong to the class of spatial rotations.
	Thus, we can redefine $\mathbf{R}_{S}\mapsto\mathbf{R}_{S}/\det(\mathbf{R}_{S})$ for $S=A,B$ to encounter cases in which the singular value decomposition includes reflections.
	The transition from panel (b) to (c) in Fig. \ref{fig:Supp2} depicts the result of the local rotations.
	Finally, this results in the standard form of the correlation matrix \eqref{eq:TrafoToStd} to which we can apply the exact solution of the EQP, cf. Eq. \eqref{eq:PentSol} and Fig. \ref{fig:Supp2}(d).

\subsection{State decomposition}

	The last point to be addressed concerns the normalization.
	In the standard form, we have $\hat\rho^\mathrm{(std)}=\sum_{a,b}P^\mathrm{(std)}(a,b)|a\rangle\langle a|\otimes|b\rangle\langle b|$.
	Thus, we can write
	\begin{align}
		\hat\rho=
		\left(\hat T_A\otimes\hat T_B\right)
		\hat\rho^\mathrm{(std)}
		\left(\hat T_A\otimes\hat T_B\right)^\dag,
	\end{align}
	with the local transformations $\hat T_S=(\hat U_S\hat T'_S)^{-1}$ begin defined through the boost operations and rotations for $S=A,B$.
	Thus, we can decompose
	\begin{align}
	\label{eq:expand}
	\begin{split}
		\hat\rho=&\sum_{a,b}
		\overbrace{P^\mathrm{(std)}(a,b)\langle a|\hat T_A^\dag\hat T_A|a\rangle\langle b|\hat T_B^\dag\hat T_B|b\rangle}^{=P(\overline a,\overline b)}
		\\
		&\phantom{\sum}\times
		\underbrace{\frac{\hat T_A|a\rangle\langle a|\hat T_A^\dag}{\langle a|\hat T_A^\dag\hat T_A|a\rangle}}_{=|\overline a\rangle\langle \overline a|}
		\otimes
		\underbrace{\frac{\hat T_B|b\rangle\langle b|\hat T_B^\dag}{\langle b|\hat T_B^\dag\hat T_B|b\rangle}}_{=|\overline b\rangle\langle \overline b|},
	\end{split}
	\end{align}
	which defines the EQP distribution $P(\overline a,\overline b)$ of $\hat\rho$ while using the transformed and normalized tensor-product states $|\overline a,\overline b\rangle$.

\subsection{Error estimate}

	In order to estimate the errors, we apply a standard Monte Carlo approach.
	That is, we repeat our decomposition with a large enough sample of $50\,000$ correlations matrices distributed according to a (multivariate) Gaussian distribution with the mean $\mathbf{C}$ and the standard deviation $\Delta\mathbf{C}$.
	The standard deviation of the sample of generated EQPs gives the error estimate as depicted in Fig. \ref{fig:Supp2}(d).

\subsection{Additional results}

	In this section, let us mention some additional results.
	First, the density operator ($\hat\rho=\sum_{k,l\in\{0,x,y,z\}}C_{k,l}\hat\sigma_k\otimes\hat\sigma_l/4$, using the sampled correlations $C_{k,l}$) reads
\begin{widetext}
	\begin{align}
		\label{eq:SampledRho}
		\hat\rho=&\left[
		\begin{array}{cccc}
			0.008 & 0.005+0.000 i & -0.002-0.001 i & -0.004-0.001 i \\
			0.005-0.000 i & 0.469 & -0.473-0.026 i & -0.006+0.002 i \\
			-0.002+0.001 i & -0.473+0.026 i & 0.500 & 0.014+0.004 i \\
			-0.004+0.001 i & -0.006-0.002 i & 0.014-0.004 i & 0.023 \\
		\end{array}
		\right],
	\end{align}
\end{widetext}
	where the error of each (complex) entry is upper bounded by $\pm0.003$.
	The purity of the reconstructed state $\hat\rho$ is $\mathrm{tr}(\hat\rho^2)=0.921\pm 0.007$.
	The overlap of the reconstructed state and the target state $|\psi\rangle=(|H,V\rangle-|V,H\rangle)\sqrt2$ is found to be $\langle\psi|\hat\rho|\psi\rangle=0.958\pm0.003$.

	The partial transposition criterion yields the negativity,
	\begin{align}
		\mathrm{Neg}(\hat\rho^{PT})=-0.459\pm0.004,
	\end{align}
	which is also the minimal eigenvalue of $\hat\rho^{PT}$.
	The actual eigenvalues of the state $\hat\rho$ are $[0.959\pm0.005,0.026\pm0.006,0.011\pm0.007,0.003\pm0.006]$, showing that the reconstructed state is a physical one (i.e., positive semidefinite).

\begin{table}[h]
	\caption{
		Expansion coefficients for states in Alice's and Bob's subsystems, cf. Eq. \eqref{eq:StatesAB}.
		The rows are listed according to $\overline a,\overline b\in\{\overline x_+,\overline x_-,\overline y_+,\overline y_-,\overline z_+,\overline z_-\}$.
	}\label{tab:Coeff}
	\begin{tabular}{rrrrrrr}
		\hline\hline
		$a_x\quad$ & $a_y\quad$ & $a_z\quad$ &\hspace*{5ex}& $b_x\quad$ & $b_y\quad$ & $b_z\quad$
		\\\hline
		$-0.907$ & $0.120$ & $-0.359$ && $-0.936$ & $0.083$ & $-0.342$ \\
		$0.944$ & $-0.224$ & $-0.197$ && $0.969$ & $-0.178$ & $-0.171$ \\
		$0.235$ & $-0.229$ & $-0.140$ && $0.384$ & $0.424$ & $-0.820$ \\
		$0.271$ & $-0.172$ & $-0.133$ && $0.187$ & $-0.791$ & $0.583$ \\
		$0.264$ & $0.059$ & $-0.107$ && $0.218$ & $-0.667$ & $-0.713$ \\
		$0.341$ & $0.031$ & $-0.175$ && $0.424$ & $0.768$ & $0.480$ \\
		\hline\hline
	\end{tabular}
\end{table}

	Finally, the vectors for both systems as used in Eq. \eqref{eq:expand} can be expanded in the following form:
	\begin{subequations}
	\label{eq:StatesAB}
	\begin{align}
		|\overline a\rangle\langle \overline a|
		=&\frac{1}{2}\left(\hat\sigma_0+a_x\hat\sigma_x+a_y\hat\sigma_y+a_z\hat\sigma_z\right),
	\end{align}
	and
	\begin{align}
		|\overline b\rangle\langle \overline b|
		=&\frac{1}{2}\left(\hat\sigma_0+b_x\hat\sigma_x+b_y\hat\sigma_y+b_z\hat\sigma_z\right).
	\end{align}
	\end{subequations}
	The corresponding coefficients as reconstructed are provided in Table \ref{tab:Coeff}.
	Together with the determined EQP, this yields the same state as given in Eq. \eqref{eq:SampledRho}, which further confirms the decomposition of the state using our EQP.


\end{document}